\documentclass[3p,times,procedia]{elsarticle}
\usepackage{nupha_ecrc}

\volume{00}

\firstpage{1}

\journalname{Nuclear Physics A}

\runauth{}

\jid{nupha}

\jnltitlelogo{Nuclear Physics A}

\usepackage{amssymb}

\usepackage[figuresright]{rotating}
\usepackage{float}
\usepackage[section]{placeins}
\usepackage{lineno}

\begin{document}

\begin{frontmatter}



\dochead{XXVIIIth International Conference on Ultrarelativistic Nucleus-Nucleus Collisions\\ (Quark Matter 2019)}

\title{Quarkonium measurements in nucleus-nucleus collisions with ALICE}


\author{Xiaozhi Bai$^{1,2}$ on behalf of the ALICE Collaboration}
\address{1 GSI Helmholtzzentrum fur Schwerionenforschung GmbH}
\address{2 Institute of Modern Physics, Chinese Academy of Sciences}
\begin{abstract}
\par Heavy quarks are produced in the early stages of nucleus-nucleus collisions and can therefore provide important insight into the Quark--Gluon Plasma (QGP). Quarkonia are proposed as crucial probes to study the QGP. The extent of the medium modification for heavy-quark quarkonium production in heavy-ion collisions is measured in terms of a nuclear modification factor $R_{\rm AA}$, defined as the quarkonium yield in heavy-ion collisions divided by the relative quarkonium cross section  in pp collisions scaled by the nuclear overlap function. A possible path-length dependent quarkonium dissociation, as well as a contribution of (re-)generation of quarkonia from heavy quarks in the medium, would lead to an azimuthal anisotropy of quarkonium production relative to the reaction plane. In this contribution, the recent ALICE measurements of quarkonium in Pb--Pb collisions at $\sqrt{s_{\rm NN}}$ = 5.02 TeV will be discussed for both mid- and forward rapidity. The dependence of $R_{\rm AA}$ on centrality and $p_{\rm T}$ for J/$\psi$, $\Upsilon$(1S), $\Upsilon$(2S),  as well as the J/$\psi$  elliptic flow $v_{2}$ will be shown. The experimental data and the current theoretical model calculations will be also discussed.
\end{abstract}

\end{frontmatter}

\section{Introduction}\label{introduction}
Heavy quarks (charm and beauty) are an excellent probe to study the strongly interacting QGP medium created in high-energy heavy-ion collisions. They are mainly produced via initial hard partonic scattering processes and thus experience the entire evolution of the QGP. Quarkonia are bound states of heavy quarks and their corresponding anti-quarks. The color screening of the surrounding medium prevent the charm and anti-charm quarks from forming their bound states. The suppression was observed in most central heavy-ion collisions at SPS and RHIC energies \cite{PhysRevLett.98.232301}. The heavy-quark production cross section is significantly higher at LHC energies \cite{PhysRevC.94.054908}. As a consequence, a new production mechanism, (re-)generation \cite{Jpsi_pbm_nature,BRAUNMUNZINGER2000196,PhysRevC.63.054905} is expected to become sizeable with increasing energy due to the higher heavy quark densities in the QGP \cite{2017212}. In particular,  in the case of charmonium at the LHC, (re-)generation is found to be the dominant production process at low transverse momentum ($p_{\rm T}$) and for central collisions \cite{PhysRevC.63.054905,ANDRONIC2019134836,DU2015147,PhysRevC.89.054911,SHI2018399}. As both suppression and (re-)generation are caused by the presence of a colored medium, quarkonium yields are indeed a sensitive probe of deconfinement of heavy quarks in the QGP.

\section{Analysis and Results}\label{analysis}
  
\par The results presented in this contribution used the data sample collected by ALICE at $\sqrt{s_{\rm NN}}$ = 5.02 TeV in 2018 for Pb--Pb, and 2017 for proton-proton (pp) collisions, respectively.  The main detectors used in the analysis  at mid-rapidity, where quarkonia are reconstructed in the di-electron channel, are the Time Projection Chamber (TPC) \cite{Alme_2010} and the Inner Tracking System (ITS) \cite{collaboration_2010}. At forward rapidity, the muon spectrometer \cite{DAS2011223} is used to reconstruct quarkonia in the di-muon channel.
\par
\begin{figure}[H]
	\centering
	\includegraphics[width=7.2cm]{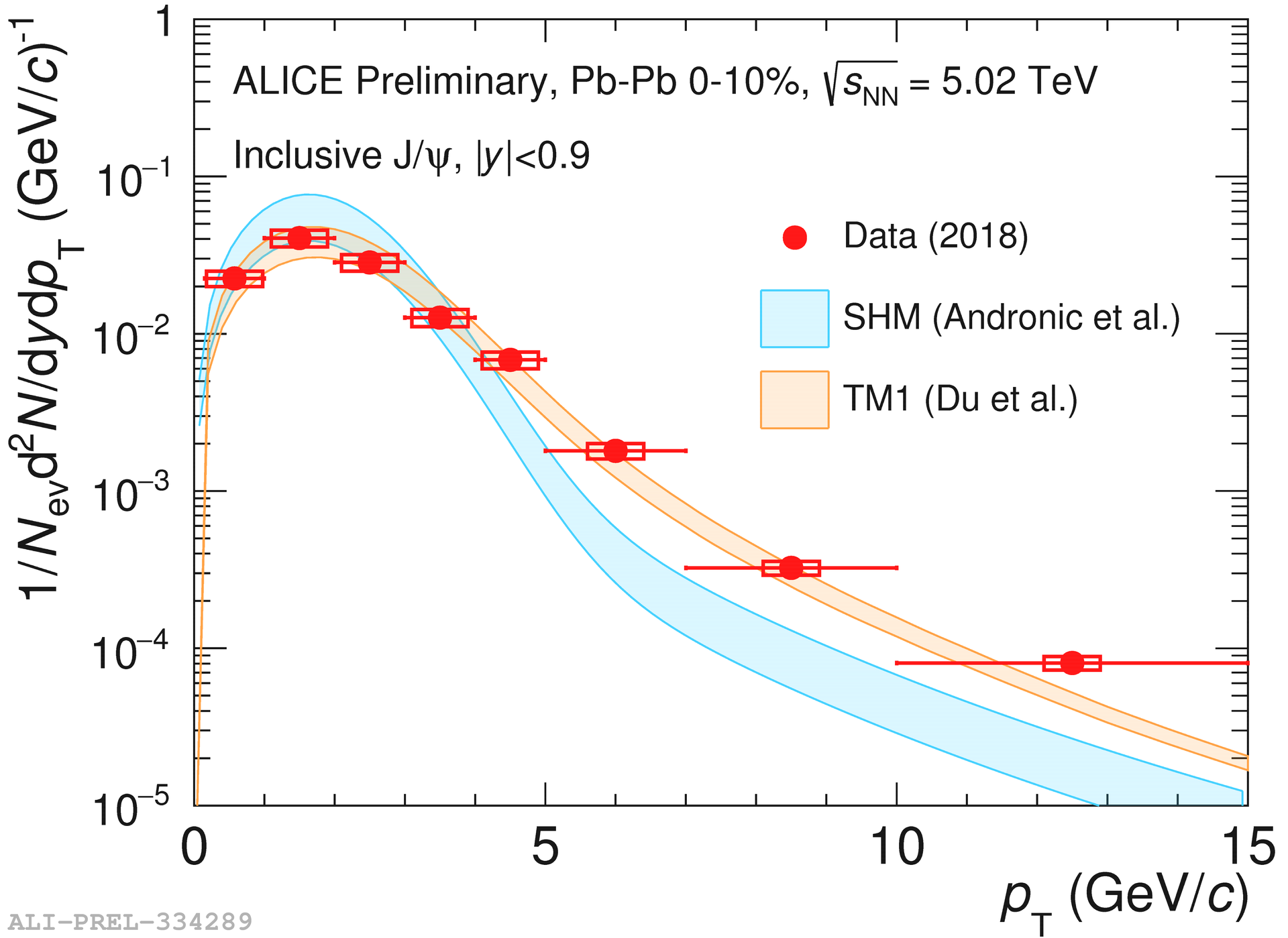}
	\includegraphics[width=7.2cm]{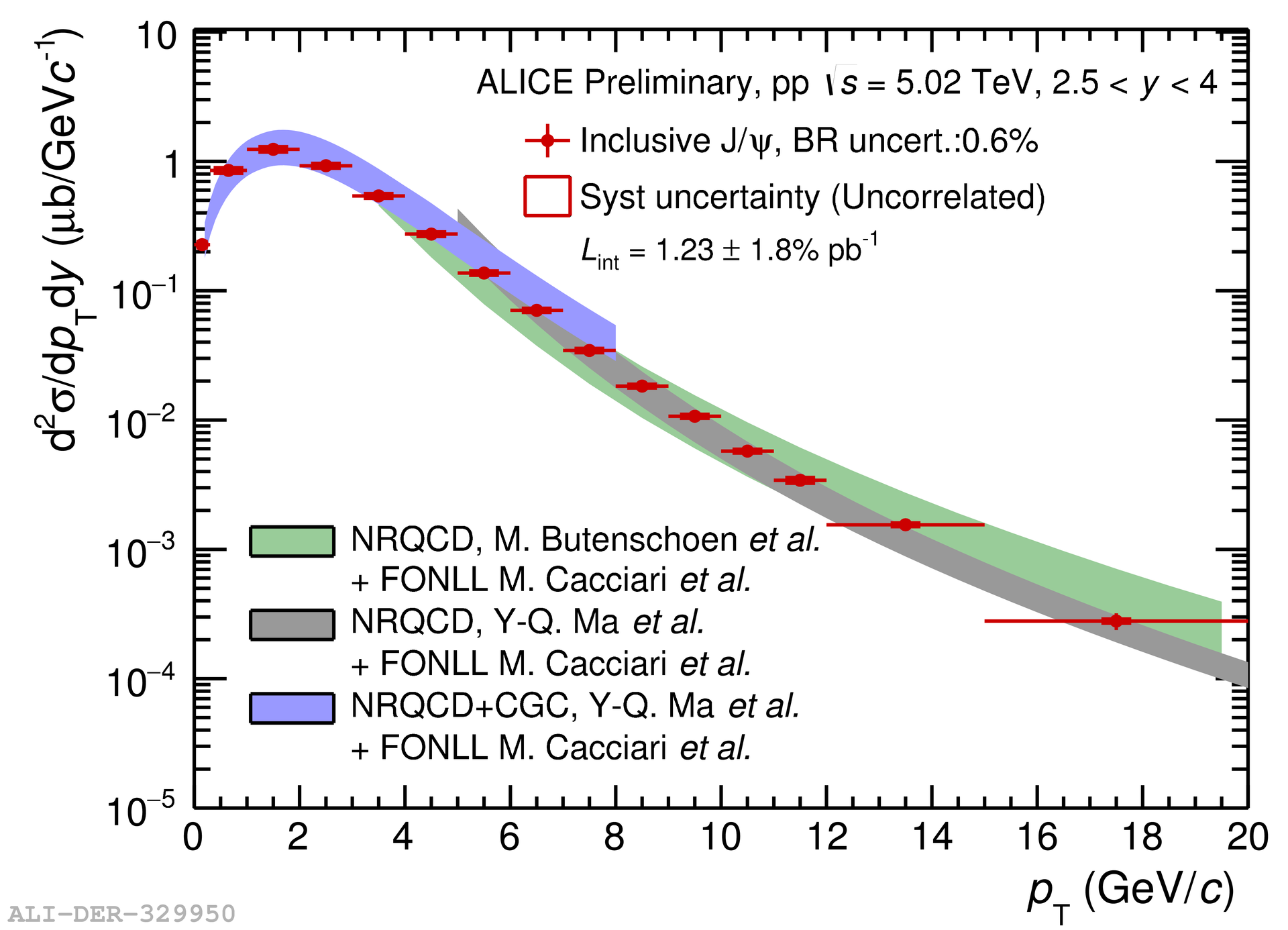}
	\caption{\textit{ Left panel:} Transverse momentum dependence of the J/$\psi$ yields in Pb--Pb collisions at $\sqrt{s_{\rm NN}}$ = 5.02 TeV at mid-rapidity for the centrality interval 0--10\%. The vertical lines and open boxes indicate the statistical and systematical uncertainties, respectively. \textit{Right panel:} The J/$\psi$ production cross section as a function of $p_{\rm T}$ at forward rapidity in pp collisions at $\sqrt{s}$ = 5.02 TeV.}
\label{fig:spectrum}
\end{figure}

 The inclusive J/$\psi$ yields as a function of the transverse momentum ($p_{\rm T}$), measured at mid-rapidity in Pb--Pb collisions at $\sqrt{s_{\rm NN}}$ = 5.02 TeV, are shown on the left panel of Fig. \ref{fig:spectrum}. The experimental data are compared to the  statistical hadronization model (SHM) \cite{ANDRONIC2019134836} and transport model (TM1) \cite{DU2015147}. The SHM describes data at low $p_{\rm T}$ only, while the TM1 agrees with data for the full $p_{\rm T}$ range.  The right panel of Fig. \ref{fig:spectrum} shows the inclusive J/$\psi$ production cross section for pp collisions at $\sqrt{s}$ = 5.02 TeV at forward rapidity. This new measurement extends the $p_{\rm T}$  coverage up to 20 GeV/$c$ and down to $p_{\rm T}$ = 0 GeV/$c$ with high precision. The data are compared to different sets of Non-Relativistic QCD (NRQCD) predictions of prompt J/$\psi$ production \cite{Butenschoen:2011ks,PhysRevD.84.114001,PhysRevLett.113.192301} combined to the contribution of non-prompt J/$\psi$ as calculated from FONLL \cite{Cacciari:1998it}. The usage of NRQCD formalism coupled to color glass condensate (CGC) \cite{PhysRevLett.113.192301} for prompt J/$\psi$ allows to extend the description of inclusive J/$\psi$ production down to $p_{\rm T}$ = 0 GeV/$c$. The agreement between all models and data is good in the whole measured $p_{\rm T}$ range.

\par
\begin{figure}[H]
	\centering
	\includegraphics[width=7.4cm]{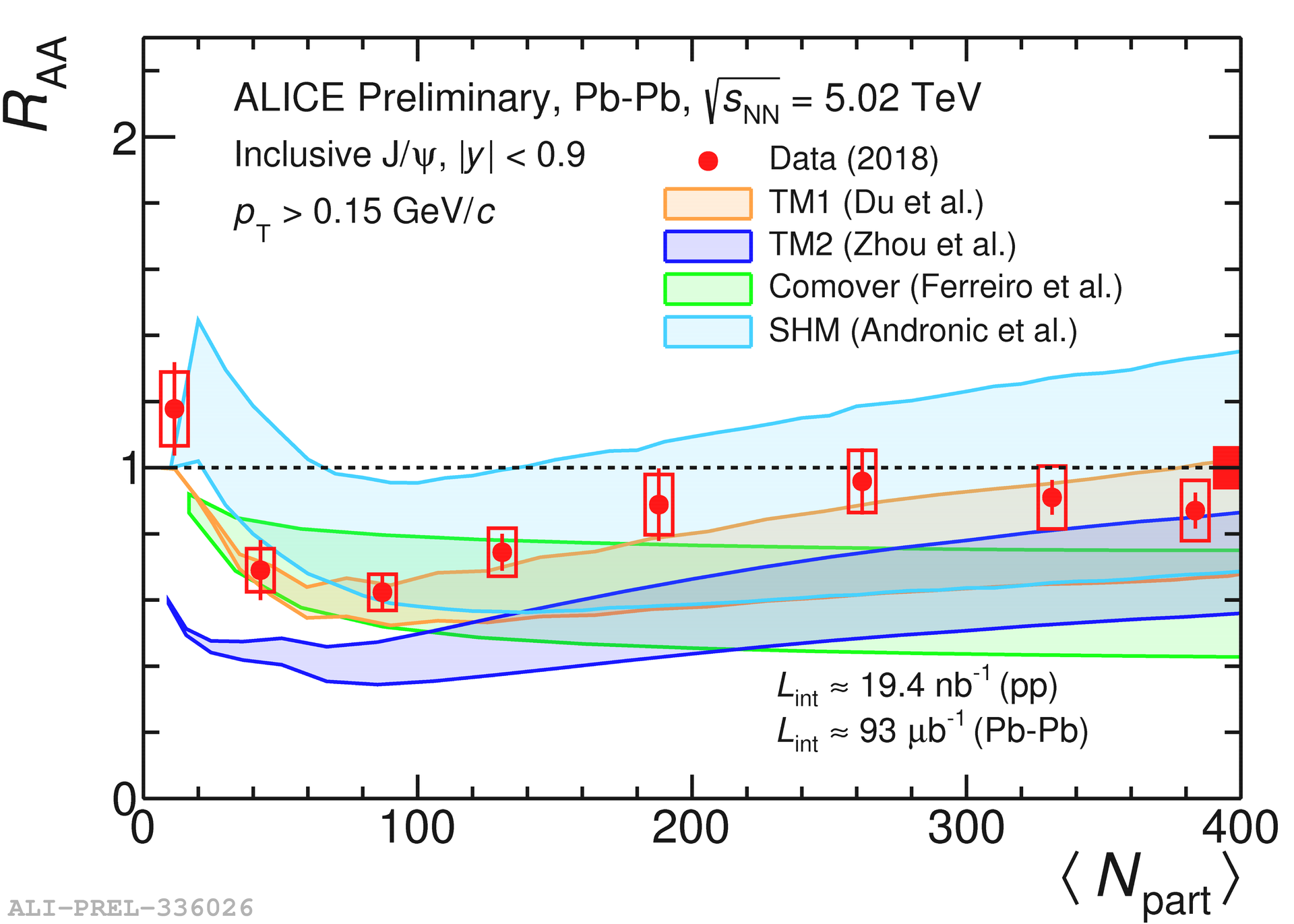}
	\includegraphics[width=7.1cm]{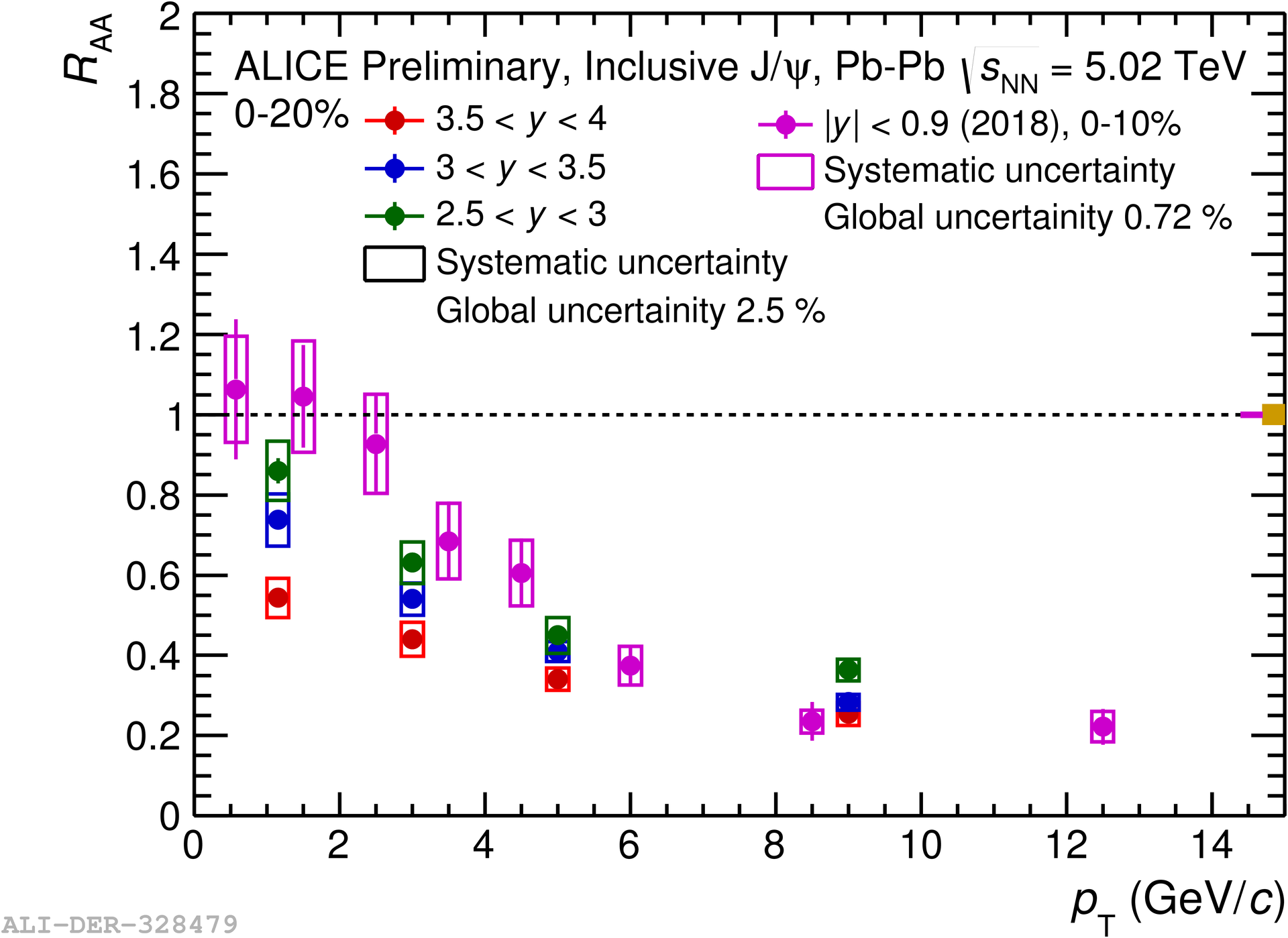}
\caption{\textit{ Left panel:} Inclusive J/$\psi$ nuclear modification factors $R_{\rm AA}$  integrated over $p_{\rm T}$ , as a function of $\langle N_{\rm part}\rangle$ in Pb--Pb collisions at $\sqrt{s_{\rm NN}}$ = 5.02 TeV at mid-rapidity. The vertical lines and open boxes indicate the statistical and systematical uncertainties, respectively.  \textit{Right panel:} The J/$\psi$ $R_{\rm AA}$ as a function of $p_{\rm T}$ in central collisions compared between different rapidity intervals. }
\label{fig:Jpsi_RAA}
\end{figure}

 The left panel of Fig. \ref{fig:Jpsi_RAA} shows the $p_{\rm T}$-integrated $R_{\rm AA}$ at mid-rapidity, in Pb--Pb collisions at $\sqrt{s_{\rm NN}}$ = 5.02 TeV. as a function of the  mean number of participants $\langle N_{\rm part}\rangle$, obtained using the pp reference for J/$\psi$ production measured at $\sqrt{s}$ = 5.02 TeV \cite{Acharya2019}. The $R_{\rm AA}$ shows an increasing trend for more central collisions ($\langle N_{\rm part}\rangle$ $>$100). The data are compared to the model calculations of SHM \cite{ANDRONIC2019134836} and TM1 \cite{DU2015147}. Both models can describe the data within uncertainties. The uncertainties of the models mainly come from the total charm quark cross section and the shadowing effects. The TM2 \cite{PhysRevC.89.054911} and comover \cite{FERREIRO201457} models underestimate the data. The right panel of Fig. \ref {fig:Jpsi_RAA}  shows the inclusive  J/$\psi$ $R_{\rm AA}$ as a function of $p_{\rm T}$ for different rapidity intervals in the centrality range 0--10\%. The data show a significant suppression at high $p_{\rm T}$ for all the rapidity intervals, 

and the suppression reduces significantly at small $p_{\rm T}$. The results also show that at low $p_{\rm T}$ the suppression reduces and even disappear when going from forward to mid-rapidity. These findings are consistent with the expectation of higher (re-)generation contribution in low $p_{\rm T}$ is higher at mid-rapidity with respect to forward rapidity.   

\begin{figure}[H]
	\centering
	\includegraphics[width=7.4cm]{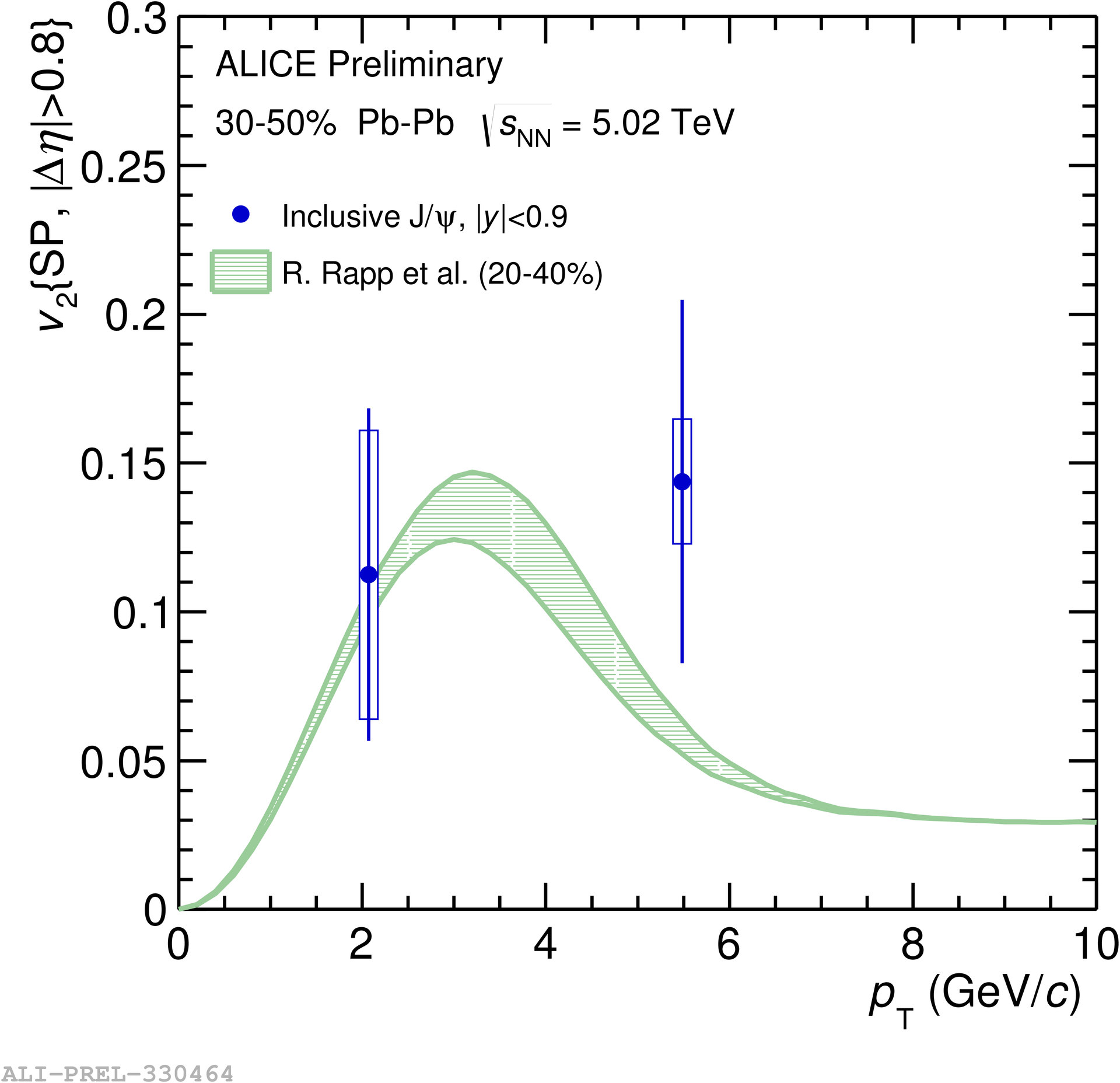}
	\includegraphics[width=7.1cm]{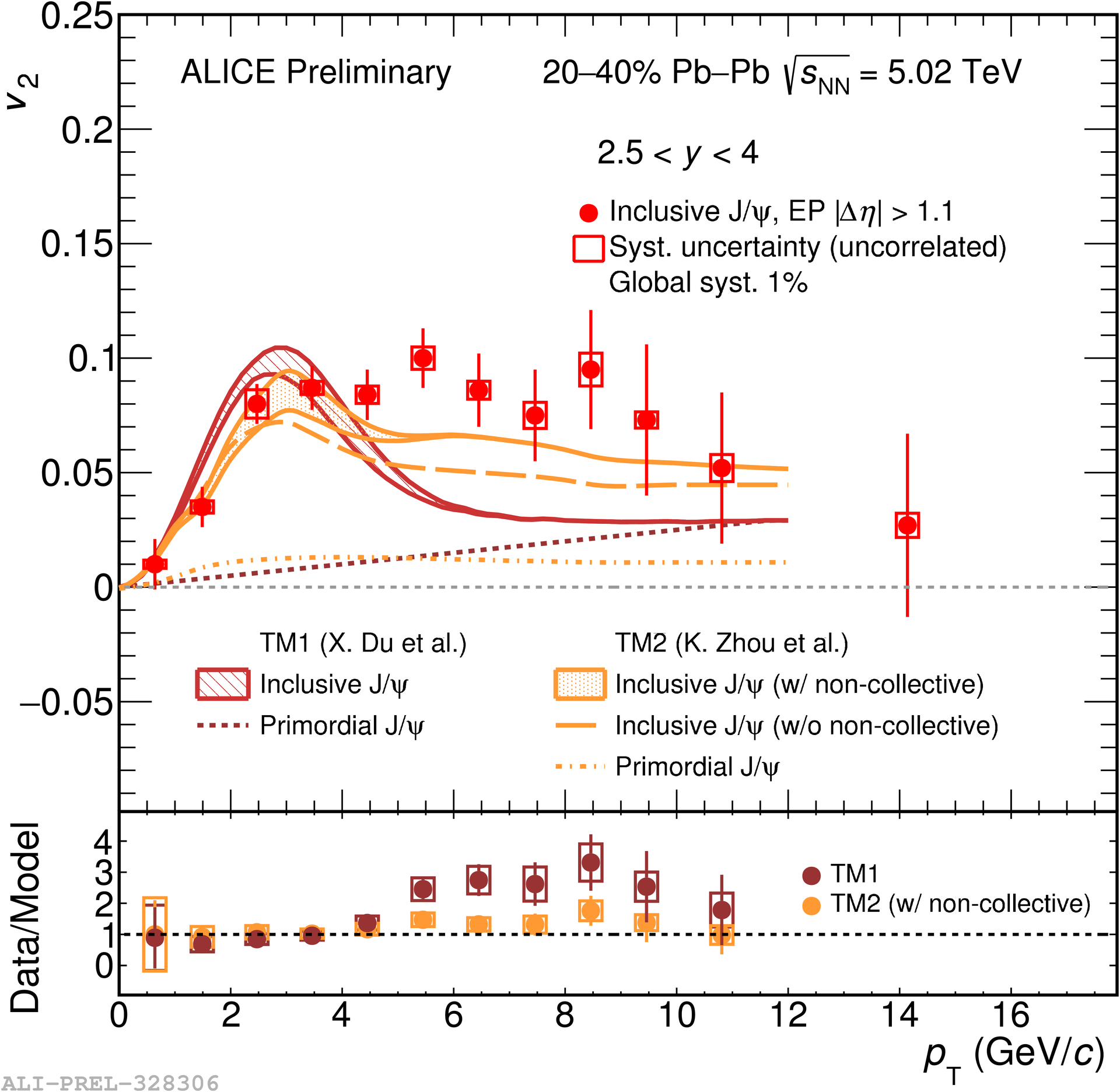}
\caption{\textit{Left panel:} Inclusive J/$\psi$ $v_{ \rm 2}$ as a function the $p_{\rm T}$ in Pb--Pb collisions at $\sqrt{s_{\rm NN}}$ = 5.02 TeV at mid-rapidity.  \textit{Right panel:} Transverse momentum dependence of the inclusive J/$\psi$ $v_{ \rm 2}$ at forward rapidity. }
\label{fig:Jpsi_V2}
\end{figure}

The inclusive $J/\psi$  elliptic flow $v_{\rm 2}$ as a function of $p_{\rm T}$ in Pb--Pb collisions at $\sqrt{s_{\rm NN}}$ = 5.02 TeV in semi-central collisions at mid-rapidity is shown on the left panel of Fig. \ref{fig:Jpsi_V2}. The data can be described by the TM1 calculation \cite{DU2015147} at low $p_{\rm T}$. The right panel of Fig. \ref{fig:Jpsi_V2}  shows the inclusive J/$\psi$ $v_{\rm 2}$ in Pb--Pb collisions at $\sqrt{s_{\rm NN}}$ = 5.02 TeV in the centrality class 20--40\% at forward rapidity, and the ratio between data and models for the TM1 and TM2. The data can be described by the TM1 calculation at low $p_{\rm T}$ ($p_{\rm T}$ $<$ 4 GeV/$c$) for both mid and forward rapidity, while they cannot successfully describe the data at high $p_{\rm T}$ ($p_{\rm T}$ $<$ 4 GeV/$c$). A positive J/$\psi$  $v_{\rm 2}$ is observed, which suggests that the charmonia are dominantly produced via the (re-)generation from thermalized charm quarks. 

\begin{figure}[H]
	\centering
	\includegraphics[width=7.4cm]{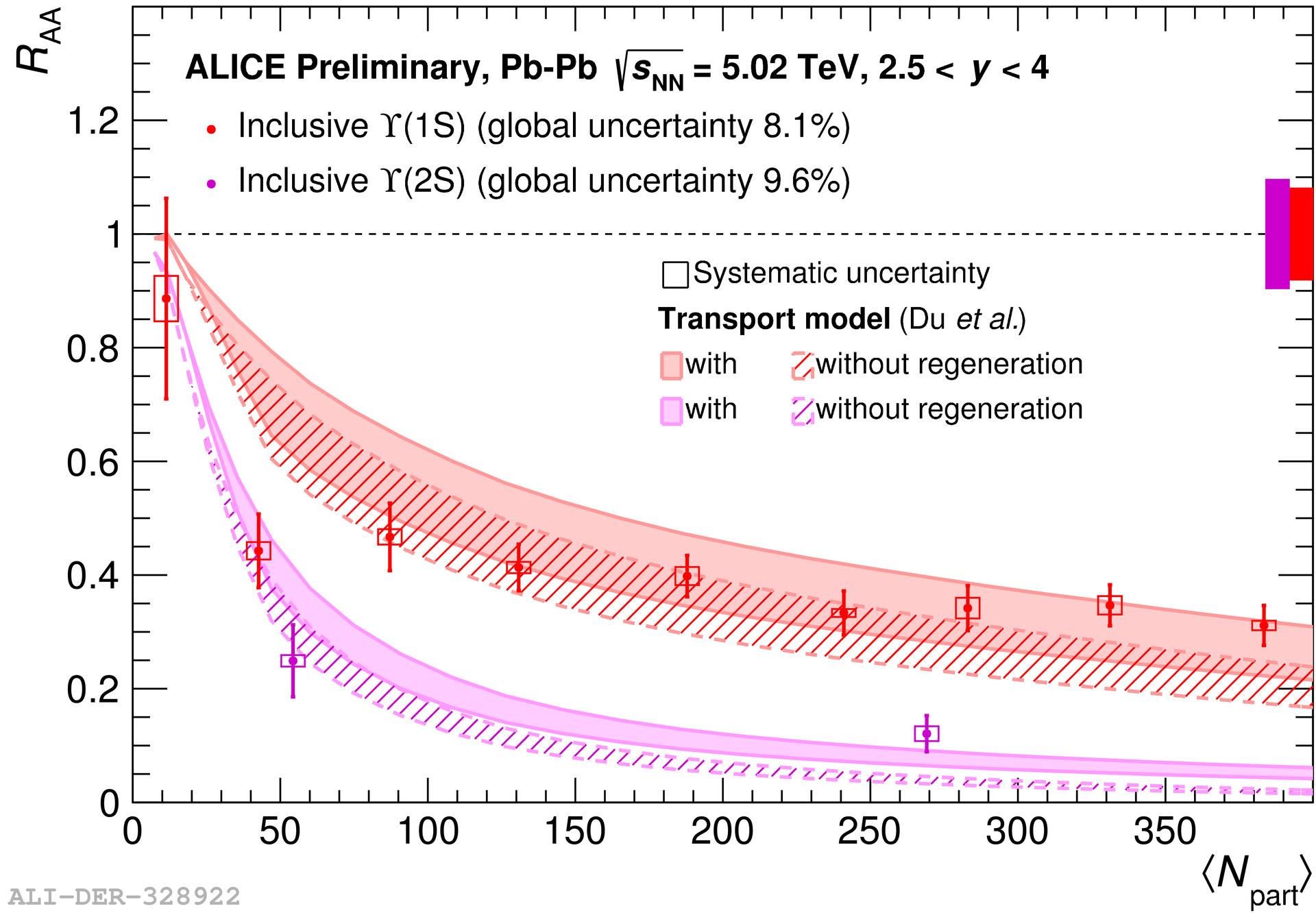}
	\includegraphics[width=7.1cm]{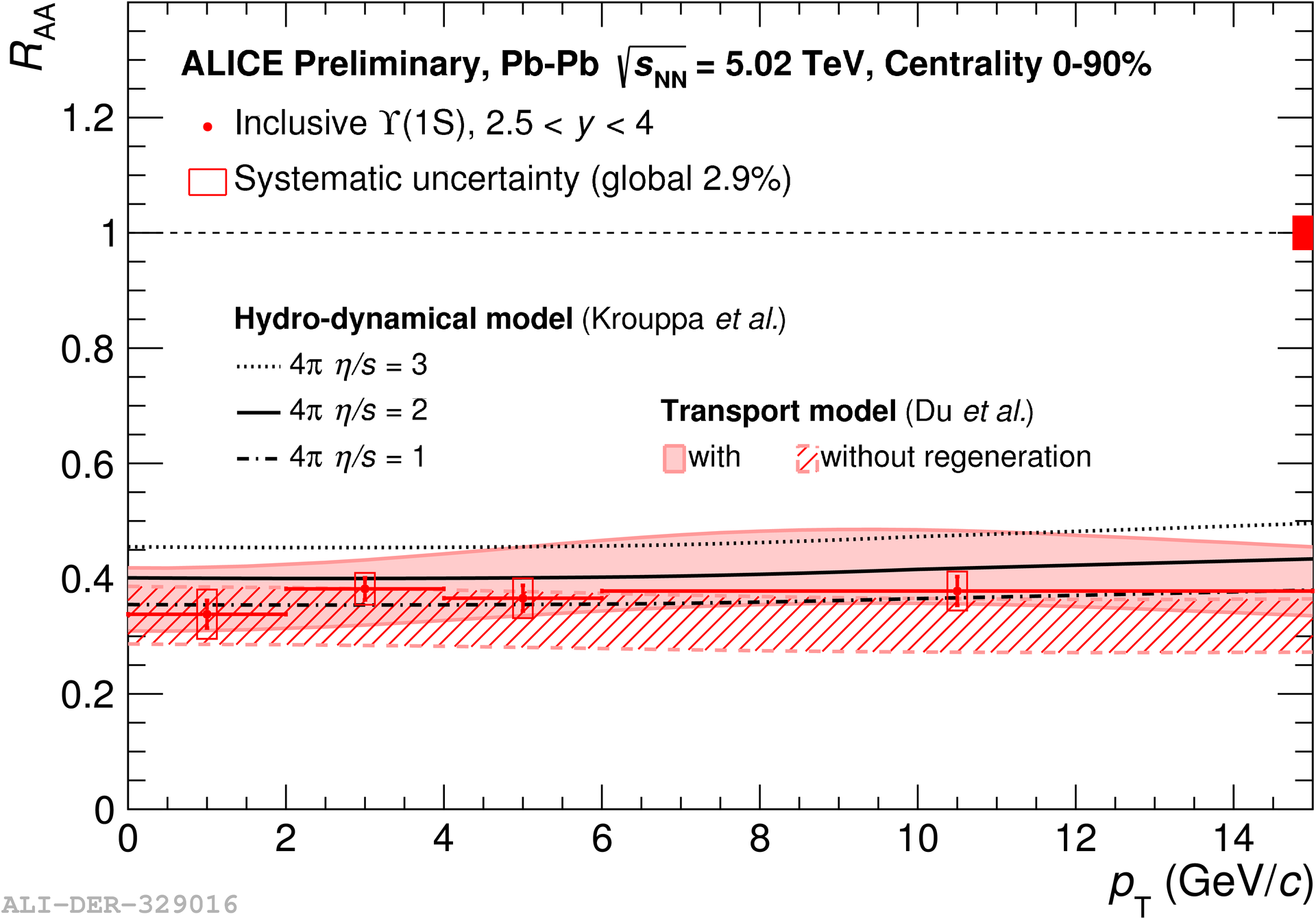}	
	\caption{\textit{ Left panel:}  The $\Upsilon$(1S) and $\Upsilon$ (2S) nuclear modification factor, integrated over $p_{\rm T}$ , as a function of the mean number of participants $\langle N_{\rm part}\rangle$ in Pb--Pb collisions at $\sqrt{s_{\rm NN}}$ = 5.02 TeV.     \textit{Right panel:} The $\Upsilon$(1S) $R_{\rm AA}$ as a function of $p_{\rm T}$.}
	\label{fig:upsilon_RAA}
\end{figure}

The  $\Upsilon$(1S) and $\Upsilon$(2S) nuclear modification factor are measured in Pb--Pb collisions at $\sqrt{s_{\rm NN}}$ = 5.02 TeV at forward rapidity using the data collected in the years 2015 and 2018. In Fig. \ref{fig:upsilon_RAA}, the left panel shows the $p_{\rm T}$-integrated  $R_{\rm AA}$ as a function of the  mean number of participants $\langle N_{\rm part}\rangle$. Both $\Upsilon$(1S) and $\Upsilon$(2S) shows only a moderate centrality dependence. $\Upsilon$(2S) shows a stronger suppression compared to $\Upsilon$(1S). The transport model \cite{PhysRevC.96.054901} agrees with data at all centralities. The precision of the data and the uncertainty of the model do not yet allow to discriminate between the cases with or without a contribution from the (re-)generation. The right panel of Fig. \ref{fig:upsilon_RAA} shows the $p_{\rm T}$ differential $R_{\rm AA}$ of the $\Upsilon$(1S) at forward rapidity. Both the hydrodynamical \cite{KROUPPA2017604} and the transport approach can describe the data within uncertainties. There is no $p_{\rm T}$ dependence, in contrast to the J/$\psi$ $R_{\rm AA}$.
\section{Summary}\label{summary}
In this contribution, the recent measurements of the quarkonium production  in Pb--Pb collisions at $\sqrt{s_{\rm NN}}$ = 5.02 TeV are discussed. The nuclear modification factor $R_{\rm AA}$ for both J/$\psi$ and $\Upsilon$ are shown as a function of the mean number of participants $\langle N_{\rm part}\rangle$ and $p_{\rm T}$. The J/$\psi$  $R_{\rm AA}$ suppression decreased gradually towards smaller rapidity, more central collisions and low $p_{\rm T}$, which indicates a dominant contribution from the (re-)generation in central collisions and low $p_{\rm T}$ for the J/$\psi$ production, while this process is negligible for bottomonium production. In fact, the $\Upsilon$ $R_{\rm AA}$ shows a strong suppression for central and semi-central collisions, and there is no significant $p_{\rm T}$ dependence. A positive J/$\psi$ $v_{\rm 2}$ is observed for both forward and mid-rapidity, which suggest the charm quarks thermalization in the QGP medium.



\bibliographystyle{elsarticle-num}
\bibliography{References.bib}

\end{document}